\documentclass[amsmath,amssymb,showkeys,superscriptaddress,floatfix,prd,10pt,aps]{revtex4-2}
\usepackage{graphicx,epstopdf}
\usepackage[caption=false]{subfig}
\usepackage{multirow}
\usepackage{appendix}
\def\pslash{p\!\!\!\slash }
\def\qslash{q\!\!\!\slash }
\def\xslash{x\!\!\!\slash }
\def\eslash{\varepsilon\!\!\!\slash }

\def\vel{\left|}
\def\ver{\right|}
\usepackage{colortbl}
\usepackage{xcolor}
\usepackage[colorlinks=true, urlcolor=blue, linkcolor=blue, citecolor=green]{hyperref}

\begin{document}

\title{Electromagnetic properties of doubly heavy pentaquark states}

\author{Ula\c{s} \"{O}zdem}%
\email[]{ulasozdem@aydin.edu.tr}
\affiliation{Health Services Vocational School of Higher Education, Istanbul Aydin University, Sefakoy-Kucukcekmece, 34295 Istanbul, Turkey}

 
\begin{abstract}
Motivated by the latest discovery of doubly-charmed tetraquark $T^+_{cc}$ by LHCb collaboration, 
we have studied the magnetic moments  of the possible doubly-heavy pentaquark states with quantum numbers $J^P = 1/2^-$ and $J^P = 3/2^-$ within the light-cone sum rules method. In the analysis, these possible pentaquark states are considered in diquark-diquark-antiquark structure. The magnetic moments of hadrons encode helpful details about the distributions of the charge and magnetization inside the hadrons, which help us to figure out their geometric configurations.
As a by product, the electric quadrupole and magnetic octupole moments of the spin-3/2 doubly-heavy pentaquark states are also extracted. These values show a non-spherical charge distribution.
It will be interesting and useful to examine the magnetic moments of these possible doubly-heavy pentaquark states with different theoretical approaches.
\end{abstract}
\keywords{Doubly-heavy pentaquarks, magnetic moment, diquark-diquark-antiquark picture, light-cone sum rule}

\maketitle

\section{Introduction}\label{motivation}

The first experimental discovery came in 2003 with the discovery of the X(3872) state \cite{Belle:2003nnu}, although they were theoretically predicted long ago that states other than conventional hadrons could exist.
Since then, scientist have paid more and more attention to the study of exotic states that are very different from conventional hadrons. 
%
%
The investigation of exotic states and how the quarks are got together inside plays a important role for comprehension the low energy QCD, and it is very crucial to search for them in experiments.
To date, many of multi-quark states have been observed by different experimental facilities.
%
Recent progress regarding these multi-quark states can be seen in Refs.
~\cite{Faccini:2012pj,Esposito:2014rxa,Esposito:2016noz,Olsen:2017bmm,Lebed:2016hpi,Nielsen:2009uh,Brambilla:2019esw, Agaev:2020zad,Chen:2016qju,Ali:2017jda,Guo:2017jvc,Liu:2019zoy, Dong:2021juy,Dong:2021bvy,Yang:2020atz}.

 In 2015,  the LHCb collaboration explored the $\Lambda_b^0$ $\rightarrow$ $J/\psi\,K^-\,p$ process and announced the observation of the two candidates for pentaquark states, $P_{c }(4380)$ and $P_{c }(4450)$, in the  $J/\psi\,p$ invariant mass distribution~\cite{Aaij:2015tga}.
In 2019, three new pentaquark states, $P_{c}(4312)$, $P_{c}(4440)$, and $P_{c}(4457)$, were reported in the updated analyses of the LHCb collaboration~\cite{Aaij:2019vzc}. 
These pentaquark states were  discovered in the $J/\psi\,p$ invariant mass spectrum, indicating that all the states include a combination of the $uudc\bar c$ quark flavors.
In 2020, the LHCb Collaboration announced a pentaquark state, $P_{cs}(4459)$, in the invariant mass spectrum of $J/\psi\Lambda$ in the $\Xi_b^0 \rightarrow J/\psi\,\Lambda\,K^-$ decay~\cite{Aaij:2020gdg}. Since this pentaquark state is observed in the invariant mass distribution $J/\psi\Lambda$, the quark content is considered to be $udsc\bar c$.
It should be noted here that most of the heavy multi-quark states discovered in the experiment so far have a hidden-charm/bottom quark structure. 
Recently, the first two charm quark-containing exotic state, $T_{cc}^+$,  was observed by the LHCb collaboration in the $D^0 D^0 \pi^+$ mass spectrum~\cite{LHCb:2021vvq,LHCb:2021auc}. Its mass,  with respect to the $D^0  D^{\ast +}$ threshold, and width have been measured to be 
%
$\delta m=m_{T_{cc}^+}-(m_{D^0}+m_{D^{\ast +}})
                   =-273 \pm 61 \pm 5_{-14}^{+11}~\mathrm{KeV}$, 
$\Gamma =410 \pm 165 \pm 43_{-38}^{+18}~\mathrm{KeV}$. 
The spin-parity of $T_{cc}^+$ tetraquark state was estimated by the experiment as $J^P = 1^+$.
Clearly, the discovery of $T^+_{cc}$ tetraquark state will open a new window to search for the new states beyond the standard hadrons, both experimentally and theoretically.
%
If the $T^+_{cc}$ is the doubly-charmed tetraquark state exist, there may also exist the doubly-charmed pentaquarks state. This argument is very similar to the underlying reasoning of predicting the hidden-charm pentaquarks from the existence of the hidden-charm tetraquark states. The only difference is that now we have two charm quarks instead of a $\bar c c$ pair. If these doubly-charmed pentaquark states do not exist, it also is important, in our opinion, to explore the reasons why they are not. Inspired by this, it is well-motivated and very interesting to search for the possible doubly-charmed pentaquark states. 
We would also like to point out that in Refs. \cite{Yang:2020twg,Chen:2021htr,Chen:2021kad}, mass and possible decay channels of possible doubly-charmed pentaquark states have been investigated within molecular pictures. In these studies, it has been shown that these possible pentaquark states are below the threshold of the possible meson-baryon state. In the study conducted in Ref. \cite{Wang:2018lhz}, possible pentaquark states are considered as compact pentaquarks and the results obtained seem to be consistent with the results obtained with the molecular pictures. Based on results of these studies, it can be seen that these possible pentaquark states might be below threshold and hence stable.

Compared to the doubly-heavy baryons, the doubly-heavy pentaquark states should be expected to be heavier. But, the sophisticated interactions within multi-quark states may lower the mass, which likely makes it hard to separate experimentally a traditional baryon from a pentaquark state just from the mass consideration. Examining other properties of these states along with mass, such as electromagnetic properties, can help shed light on the inner nature of these states.
The magnetic dipole and higher moments of hadrons can help us to obtain helpful details on the charge and magnetization distributions as well as their geometric shape. 
In this work,  we study the magnetic moments of spin-1/2 and spin-3/2 doubly-heavy pentaquark (For short $P^{1/2}_{QQ}$ and $P^{3/2}_{QQ}$, respectively) states using the light-cone sum rule formalism~\cite{Chernyak:1990ag, Braun:1988qv, Balitsky:1989ry} in the compact pentaquark picture. The attractive interaction induced by one-gluon exchange supports formation of the diquarks in color antitriplet and the supported configurations are the scalar and axialvector diquark states from the QCD sum rules \cite{Jamin:1989hh,Wang:2010sh,Tang:2012np,Kleiv:2013dta}. In case of the heavy diquark systems, only the tensor and axialvector diquarks remain because of the Fermi-Dirac statistics, the axialvector diquarks are more stable than the tensor diquarks.  Therefore, in this study, we choose the axialvector type heavy diquark interpolating currents.   
There are some theoretical estimations on the internal structure of $P^{1/2}_{QQ}$ and $P^{3/2}_{QQ}$  states, their masses, production mechanisms and decay channels using different configurations and models~\cite{Yang:2020twg,Chen:2021htr,Chen:2021kad,Wang:2018lhz,Zhou:2018bkn,Park:2018oib,Xu:2010fc,Chen:2017vai,Shimizu:2017xrg}.
It should be noted here that, magnetic moments of hidden-charm pentaquark states have been obtained via different models and substructures~\cite{Wang:2016dzu, Ortiz-Pacheco:2018ccl,Xu:2020flp,Ozdem:2021btf,Ozdem:2018qeh,Ozdem:2021ugy, Li:2021ryu,Gao:2021hmv}. 
 Though the short lifetimes of the pentaquark states make the magnetic moment difficult to be measured at present, more data accumulation in different experiments in the future may make this possible. 
 The $\Delta^+(1232)$ baryon has also a very short lifetime, however, its magnetic moment have been obtained by means of $\gamma N $ $ \rightarrow $ $ \Delta $ $\rightarrow $ $ \Delta \gamma $ $ \rightarrow$ $ \pi N \gamma $ process~ \cite{Pascalutsa:2004je, Pascalutsa:2005vq, Pascalutsa:2007wb}. 
The electromagnetic properties of baryons containing two charm quarks have been calculated in the framework of the lattice QCD method~\cite{Can:2013zpa,Can:2013tna}, and it may be possible to generalize these analyzes to exotic states in the near future.

This work has the following structure.  In Sect. \ref{formalism}, we briefly discuss the formalism and calculate the light-cone sum rule for the magnetic moments under investigation. In Sect. \ref{numerical}, the numerical analysis and discussions for the magnetic moments are presented. A brief summary of the article is presented in Sect. \ref{summary}. 
The explicit expressions of the photon distribution amplitudes and  magnetic moments for spin-1/2 doubly-heavy pentaquark states are given in Appendixes A and B.

\begin{widetext}
 
\section{Formalism}\label{formalism}

For magnetic moment analysis, calculations are started by writing the appropriate correlation function in the light-cone sum rules. This correlation function allows us to calculate the physical quantity to be calculated, in our case, the magnetic moment, in terms of both QCD and hadron parameters. Then, the correlation function calculated in two different ways is equalized to each other using the quark-hadron duality. As a final step, Borel transform and continuum subtraction are performed to reduce continuum and higher states effects.

\subsection{Formalism of the \texorpdfstring{$P^{1/2}_{QQ}$}{} states}\label{for:Pcc12}

The correlation function required for the computations magnetic moment has the following form:
 
\begin{eqnarray} \label{edmn01}
\Pi(p,q)&=&i\int d^4x e^{ip \cdot x} \langle0|T\left\{J^{P^{1/2}_{QQ}}(x)\bar{J}^{P^{1/2}_{QQ}}(0)\right\}|0\rangle _\gamma \, ,
\end{eqnarray}
where $\gamma$ is the external electromagnetic field and the $J(x)$ stands for interpolating currents of the considered spin-1/2 doubly-heavy pentaquark states. 
We would like to point out that many possible interpolating currents can be written for exotic states, but the number of possible interpolating currents that can be written can be slightly reduced when the QCD sum rules and the states to be examined are taken into account. As we mentioned in the introduction of the text, only axialvector or tensor structures make significant contributions in case of heavy-heavy diquark configurations. Among these two structures, axialvector structures were preferred because they are more stable (Considering the heavy quark limit, it can be seen that the tensor structure is proportional to zero).  Apart from the heavy-heavy diquark case, heavy-light diquarks may also be in question. However, due to the lack of spectroscopic parameters in the QCD sum rules, they are excluded from this work for now, considering that they can be examined in future studies.
Therefore, in this work, we choose the axialvector type heavy-heavy diquark interpolating current and its given as follows~\cite{Wang:2018lhz}:
\begin{eqnarray}
  J^{P^{1/2}_{QQ}}(x)&=& \varepsilon^{abc}\varepsilon^{ade} \varepsilon^{bfg} \big[Q^T_d(x) C\gamma_\mu Q_e(x)\big]\big[ u^T_f(x) C\gamma_5 d_g(x)\big] \gamma_5\gamma^\mu C \bar{q}^T_c(x) \, ,
\end{eqnarray}
where $Q $ is c or b-quark and $q$ is u or d-quark.

In order to get the hadronic degrees of freedom of the correlation function, we insert a complete set of intermediate $P^{1/2}_{QQ}$ states with the same quantum numbers as the interpolating currents into the correlation function. As a result, we get 
 \begin{align}\label{edmn02}
\Pi^{Had}(p,q)&=\frac{\langle0\mid J(x) \mid
{P^{1/2}_{QQ}}(p, s) \rangle}{[p^{2}-m_{P^{1/2}_{QQ}}^{2}]}
\langle {P^{1/2}_{QQ}}(p, s)\mid
{P^{1/2}_{QQ}}(p+q, s)\rangle_\gamma 
\frac{\langle {P^{1/2}_{QQ}}(p+q, s)\mid
\bar{J}(0) \mid 0\rangle}{[(p+q)^{2}-m_{{P^{1/2}_{QQ}}}^{2}]}+\cdots 
\end{align}

The matrix element  $\langle
{P^{1/2}_{QQ}}(p, s)\mid {P^{1/2}_{QQ}}(p+q, s)\rangle_\gamma$ inserting Eq. (\ref{edmn02}) can be described in connection with form factors as follows:
%
\begin{align} \label{edmn04}
\langle {P^{1/2}_{QQ}}(p, s)\mid {P^{1/2}_{QQ}}(p+q, s)\rangle_\gamma &=\varepsilon^\mu\,\bar u(p, s)\Big[\big[F_1(q^2)
+F_2(q^2)\big] \gamma_\mu +F_2(q^2)
\frac{(2p+q)_\mu}{2 m_{P^{1/2}_{QQ}}}\Big]\,u(p+q, s).
\end{align}
%

We insert Eq.~(\ref{edmn04}) in Eq. (\ref{edmn02}). Then, after some calculations are made, we get the result for the hadronic side as follows
%
%
\begin{align}
\label{edmn05}
\Pi^{Had}(p,q)=&\lambda^2_{P^{1/2}_{QQ}}\gamma_5 \frac{\Big(\pslash+m_{P^{1/2}_{QQ}} \Big)}{[p^{2}-m_{{P^{1/2}_{QQ}}}^{2}]}\varepsilon^\mu \Bigg[\big[F_1(q^2) %
+F_2(q^2)\big] \gamma_\mu
+F_2(q^2)\, \frac{(2p+q)_\mu}{2 m_{P^{1/2}_{QQ}}}\Bigg]  \gamma_5 
\frac{\Big(\pslash+\qslash+m_{P^{1/2}_{QQ}}\Big)}{[(p+q)^{2}-m_{{P^{1/2}_{QQ}}}^{2}]}. 
\end{align}
The value of  form factors $F_1(q^2)$ and $F_2(q^2)$ give us the  magnetic form factor $F_M(q^2)$ at different $q^2$ :
\begin{align}
\label{edmn07}
&F_M(q^2) = F_1(q^2) + F_2(q^2).
\end{align}
 At static limit, i.e. $q^2 = 0 $,   magnetic form factor $F_M (q^2 = 0)$ is proportional to the magnetic moment $\mu_{P_{QQ}}$ for real photon :
\begin{align}
\label{edmn08}
&\mu_{P_{QQ}} = \frac{ e}{2\, m_{P_{QQ}}} \,F_M(q^2 = 0).
\end{align}

The next step is to obtain the correlation function in terms of QCD degrees of freedom.
When calculating the correlation function in terms of QCD degrees of freedom, the explicit forms of the interpolating currents are substituted into the correlation function. Then, the relevant light and heavy quark fields are contracted via the Wick's theorem, and the desired results are obtained. 
Consequently, we get 

\begin{eqnarray}\label{QCDside1}
\Pi^{QCD}(p,q)&=&-i\,\varepsilon_{abc} \varepsilon_{ade} \varepsilon_{bfg} \varepsilon_{a^{\prime}d^{\prime}e^{\prime}} 
\varepsilon_{a^{\prime}b^{\prime}c^{\prime}} \varepsilon_{b^{\prime}f^{\prime}g^{\prime}} \int d^4x \, e^{ip\cdot x} \langle 0\mid \gamma_5 \gamma^\mu \widetilde S_q^{c^{\prime}c}(-x) \gamma^\nu \gamma_5 
\nonumber\\
&& \Big\{
\mbox{Tr}\Big[\gamma_\mu S_{Q}^{ee^\prime}(x) \gamma_\nu \widetilde S_{Q}^{dd^\prime}(x)\Big]\, \mbox{Tr}\Big[\gamma_5 S_d^{gg^\prime}(x)  \gamma_5 \widetilde S_{u}^{ff^\prime}(x)\Big]  \nonumber\\ 
&& 
- 
\mbox{Tr}\Big[\gamma_\mu S_{Q}^{ed^\prime}(x) \gamma_\nu \widetilde S_{Q}^{de^\prime}(x)\Big]\, \mbox{Tr}\Big[\gamma_5 S_d^{gg^\prime}(x)  \gamma_5 \widetilde S_{u}^{ff^\prime}(x)\Big]\Big\}   \mid 0 \rangle _\gamma \, ,
\end{eqnarray}
where
$\widetilde{S}_{Q(q)}^{ij}(x)=CS_{Q(q)}^{ij\mathrm{T}}(x)C$ and ,
 $S_{q}(x)$ and $S_{Q}(x)$ are the light and heavy quark propagators, respectively. Their explicit expressions in the x-space are presented as
\begin{align}
\label{edmn13}
S_{q}(x)&=i \frac{{\xslash}}{2\pi ^{2}x^{4}} 
- \frac{\langle \bar qq \rangle }{12} \Big(1-i\frac{m_{q} \xslash}{4}   \Big)
- \frac{ \langle \bar qq \rangle }{192}m_0^2 x^2  \Big(1
-i\frac{m_{q} \xslash}{6}   \Big)
-\frac {i g_s }{32 \pi^2 x^2} ~G^{\mu \nu} (x) \Big[\rlap/{x} 
\sigma_{\mu \nu} +  \sigma_{\mu \nu} \rlap/{x}
 \Big],\\
\nonumber\\
\label{edmn14}
S_{Q}(x)&=\frac{m_{Q}^{2}}{4 \pi^{2}} \Bigg[ \frac{K_{1}\Big(m_{Q}\sqrt{-x^{2}}\Big) }{\sqrt{-x^{2}}}
+i\frac{{\xslash}~K_{2}\Big( m_{Q}\sqrt{-x^{2}}\Big)}
{(\sqrt{-x^{2}})^{2}}\Bigg]
-\frac{g_{s}m_{Q}}{16\pi ^{2}} \int_0^1 dv\, G^{\mu \nu }(vx)\Bigg[ (\sigma _{\mu \nu }{\xslash}
  +{\xslash}\sigma _{\mu \nu })
  \nonumber\\
  &\times \frac{K_{1}\Big( m_{Q}\sqrt{-x^{2}}\Big) }{\sqrt{-x^{2}}}
+2\sigma_{\mu \nu }K_{0}\Big( m_{Q}\sqrt{-x^{2}}\Big)\Bigg].
\end{align}%
where $\langle \bar qq \rangle$ is quark  condensate, $m_0$ is characterized via the quark-gluon mixed condensate  $\langle 0 \mid \bar  q\, g_s\, \sigma_{\mu\nu}\, G^{\mu\nu}\, q \mid 0 \rangle = m_0^2 \,\langle \bar qq \rangle $, $K_0$, $K_1$ and $K_2$ are modified the second kind Bessel functions, $v$ is line variable and $G^{\mu\nu}$ is the gluon field strength tensor. 
 The first term of the light and massive quark propagators corresponds to the free or perturbative part, and the remaining part related to the interacting parts.

The correlation function in Eq. (\ref{QCDside1}) includes  different types of contributions: The photon can be emitted both perturbatively and
non-perturbatively. 
In the first case, one of the free light or massive quark 
propagators in Eq.~(\ref{QCDside1}) is replaced by
\begin{align}
\label{sfree}
S^{free} \rightarrow \int d^4y\, S^{free} (x-y)\,\rlap/{\!A}(y)\, S^{free} (y)\,,
\end{align}
the remaining four propagators are replaced with the full quark propagators.
The light-cone sum rule calculations are most conveniently done in the fixed-point gauge. For electromagnetic field,
it is defined by $x_\mu A^\mu =0$. 
In this gauge, the electromagnetic potential is given by
\begin{align}
\label{AAA}
 &A_\alpha = -\frac{1}{2} F_{\alpha\beta}y^\beta 
   = -\frac{1}{2} (\varepsilon_\alpha q_\beta-\varepsilon_\beta q_\alpha)\,y^\beta.
\end{align}
Equation (\ref{AAA}) is plugged into Eq. (\ref{sfree}), as a result of which we obtain
 \begin{align}
  S^{free} \rightarrow -\frac{1}{2} (\varepsilon_\alpha q_\beta-\varepsilon_\beta q_\alpha)
  \int\, d^4y \,y^{\beta}\, 
  S^{free} (x-y)\,\gamma_{\alpha}\,S^{free} (y)\,,
 \end{align}

After some calculations for $S_q^{free}$ and $S_Q^{free}$  we get

\begin{eqnarray}
&& S_q^{free}=\frac{e_q}{32 \pi^2 x^2}\Big(\varepsilon_\alpha q_\beta-\varepsilon_\beta q_\alpha\Big)
 \Big(\xslash\sigma_{\alpha \beta}+\sigma_{\alpha\beta}\xslash\Big),\nonumber\\
&& S_Q^{free}=-i\frac{e_Q m_Q}{32 \pi^2}
\Big(\varepsilon_\alpha q_\beta-\varepsilon_\beta q_\alpha\Big)
\Big[2\sigma_{\alpha\beta}K_{0}\Big( m_{Q}\sqrt{-x^{2}}\Big)
 +\frac{K_{1}\Big( m_{Q}\sqrt{-x^{2}}\Big) }{\sqrt{-x^{2}}}
 \Big(\xslash\sigma_{\alpha \beta}+\sigma_{\alpha\beta}\xslash\Big)\Big].
\end{eqnarray}

In the second case one of the light quark 
propagators in Eq.~(\ref{QCDside1}) are replaced by
\begin{align}
\label{neweq}
S_{\alpha\beta}^{ab} \rightarrow -\frac{1}{4} (\bar{q}^a \Gamma_i q^b)(\Gamma_i)_{\alpha\beta},
\end{align}
and the remaining propagators are full quark propagators including 
the perturbative as well as the nonperturbative contributions.
Here as an example, we give a short detail of the calculations of the QCD representations.
 In second case for simplicity, we only consider the first trace in Eq.~(\ref{QCDside1}),
\begin{align}
\label{QCDES}
\Pi^{QCD}(p,q)&=-i\,\varepsilon_{abc} \varepsilon_{ade} \varepsilon_{bfg} \varepsilon_{a^{\prime}d^{\prime}e^{\prime}} 
\varepsilon_{a^{\prime}b^{\prime}c^{\prime}} \varepsilon_{b^{\prime}f^{\prime}g^{\prime}} \int d^4x \, e^{ip\cdot x} \langle 0\mid \gamma_5 \gamma^\mu \widetilde S_q^{c^{\prime}c}(-x) \gamma^\nu \gamma_5 
\nonumber\\
& \Big\{
\mbox{Tr}\Big[\gamma_\mu S_{Q}^{ee^\prime}(x) \gamma_\nu \widetilde S_{Q}^{dd^\prime}(x)\Big]\, \mbox{Tr}\Big[\gamma_5 S_d^{gg^\prime}(x)  \gamma_5 \widetilde S_{u}^{ff^\prime}(x)\Big] 
\Big\}   \mid 0 \rangle _\gamma \, +...
 \end{align}
 
 By replacing one of light propagators with the expressions in Eq. (\ref{edmn13})
 and making use of 
 \begin{align}
  \bar q^a(x)\Gamma_i q^{a'}(0)\rightarrow \frac{1}{3}\delta^{aa'}\bar q(x)\Gamma_i q(0),
 \end{align}
the Eq. (\ref{QCDES}) takes the form
 \begin{align}
\label{QCDES2}
\Pi^{QCD}(p,q)&=i\,\varepsilon_{abc} \varepsilon_{ade} \varepsilon_{bfg} \varepsilon_{a^{\prime}d^{\prime}e^{\prime}} 
\varepsilon_{a^{\prime}b^{\prime}c^{\prime}} \varepsilon_{b^{\prime}f^{\prime}g^{\prime}} \int d^4x \, e^{ip\cdot x}  
\nonumber\\
&\Bigg\{ \gamma_5 \gamma^\mu \Gamma_i \gamma^\nu \gamma_5 
\mbox{Tr}\Big[\gamma_\mu S_{Q}^{ee^\prime}(x) \gamma_\nu \widetilde S_{Q}^{dd^\prime}(x)\Big]\, \mbox{Tr}\Big[\gamma_5 S_d^{gg^\prime}(x)  \gamma_5 \widetilde S_{u}^{ff^\prime}(x)\Big]\delta^{c'c} \nonumber\\
&+ \gamma_5 \gamma^\mu \widetilde S_q^{c^{\prime}c}(-x) \gamma^\nu \gamma_5 
\mbox{Tr}\Big[\gamma_\mu S_{Q}^{ee^\prime}(x) \gamma_\nu \widetilde S_{Q}^{dd^\prime}(x)\Big]\, \mbox{Tr}\Big[\gamma_5\Gamma_i  \gamma_5 \widetilde S_{u}^{ff^\prime}(x)\Big]\delta^{gg'}\nonumber\\
&+
 \gamma_5 \gamma^\mu \widetilde S_q^{c^{\prime}c}(-x) \gamma^\nu \gamma_5 
\mbox{Tr}\Big[\gamma_\mu S_{Q}^{ee^\prime}(x) \gamma_\nu \widetilde S_{Q}^{dd^\prime}(x)\Big]\, \mbox{Tr}\Big[\gamma_5 S_d^{gg^\prime}(x)  \gamma_5 \Gamma_i \Big] \delta^{ff'}
\Bigg\} \frac{1}{12} \langle \gamma(q) |\bar q(x)\Gamma_i q(0)|0\rangle 
+...,
 \end{align}
where $\Gamma_i = I, \gamma_5, \gamma_\mu, i\gamma_5 \gamma_\mu, \sigma_{\mu\nu}/2$.
Similarly, when a light propagator interacts with the photon, 
a gluon may be released from one of the remaining four propagators. 
The expression obtained in this case is as follows:
\begin{align}
\label{QCDES3}
 \Pi _{\mu \nu }^{\mathrm{QCD}}(p,q)&=i\,\varepsilon_{abc} \varepsilon_{ade} \varepsilon_{bfg} \varepsilon_{a^{\prime}d^{\prime}e^{\prime}} 
\varepsilon_{a^{\prime}b^{\prime}c^{\prime}} \varepsilon_{b^{\prime}f^{\prime}g^{\prime}} \int d^4x \, e^{ip\cdot x}  
\nonumber\\
&\Bigg\{ \gamma_5 \gamma^\mu \Gamma_i \gamma^\nu \gamma_5 
\mbox{Tr}\Big[\gamma_\mu S_{Q}^{ee^\prime}(x) \gamma_\nu \widetilde S_{Q}^{dd^\prime}(x)\Big]\, \mbox{Tr}\Big[\gamma_5 S_d^{gg^\prime}(x)  \gamma_5 \widetilde S_{u}^{ff^\prime}(x)\Big] 
\Big[\Big(\delta^{ce}\delta^{c'e'}  -\frac{1}{3}\delta^{cc'}\delta^{ee'}\Big)\nonumber\\
&+\Big(\delta^{cd}\delta^{c'd'}  -\frac{1}{3}\delta^{cc'}\delta^{dd'}\Big) + 
\Big(\delta^{cg}\delta^{c'g'}  -\frac{1}{3}\delta^{cc'}\delta^{gg'}\Big) +
\Big(\delta^{cf}\delta^{c'f'}  -\frac{1}{3}\delta^{cc'}\delta^{ff'}\Big)
\Big]\nonumber\\
&+\gamma_5 \gamma^\mu \widetilde S_q^{c^{\prime}c}(-x) \gamma^\nu \gamma_5 
\mbox{Tr}\Big[\gamma_\mu S_{Q}^{ee^\prime}(x) \gamma_\nu \widetilde S_{Q}^{dd^\prime}(x)\Big]\, \mbox{Tr}\Big[\gamma_5\Gamma_i  \gamma_5 \widetilde S_{u}^{ff^\prime}(x)\Big]\Big[\Big(\delta^{gc}\delta^{g'c'}  -\frac{1}{3}\delta^{gg'}\delta^{cc'}\Big)\nonumber\\
&+\Big(\delta^{ge}\delta^{g'e'}  -\frac{1}{3}\delta^{ee'}\delta^{gg'}\Big)+ \Big(\delta^{gd}\delta^{g'd'}  -\frac{1}{3}\delta^{gg'}\delta^{dd'}\Big) +
\Big(\delta^{gf}\delta^{g'f'}  -\frac{1}{3}\delta^{gg'}\delta^{ff'}\Big)\Big]\nonumber\\
&+\gamma_5 \gamma^\mu \widetilde S_q^{c^{\prime}c}(-x) \gamma^\nu \gamma_5 
\mbox{Tr}\Big[\gamma_\mu S_{Q}^{ee^\prime}(x) \gamma_\nu \widetilde S_{Q}^{dd^\prime}(x)\Big]\, \mbox{Tr}\Big[\gamma_5 S_d^{gg^\prime}(x)  \gamma_5 \Gamma_i \Big]\Big[\Big(\delta^{fc}\delta^{f'c'}  -\frac{1}{3}\delta^{ff'}\delta^{cc'}\Big)\nonumber\\
&+ \Big(\delta^{fe}\delta^{f'e'}  -\frac{1}{3}\delta^{ff'}\delta^{ee'}\Big)+ \Big(\delta^{fd}\delta^{f'd'}  -\frac{1}{3}\delta^{ff'}\delta^{dd'}\Big) + \Big(\delta^{fg}\delta^{f'g'}  -\frac{1}{3}\delta^{ff'}\delta^{gg'}\Big)\Big]
\Bigg\}
\nonumber\\
& \times \frac{1}{32} \langle \gamma(q) |\bar q(x)\Gamma_i G_{\mu\nu}(vx) q(0)|0\rangle 
+...,
 \end{align}
where we inserted
\begin{align}
\label{QCDES5}
 \bar q^a(x)\Gamma_i G_{\mu\nu}^{bb'}(vx) q^{a'}(0)\rightarrow \frac{1}{8}\Big(\delta^{ab}\delta^{a'b'}
 -\frac{1}{3}\delta^{aa'}\delta^{bb'}\Big)\bar q(x)\Gamma_i G_{\mu\nu}(vx) q(0).
\end{align}

As is seen, there appear matrix
elements such as $\langle \gamma(q)\vel \bar{q}(x) \Gamma_i q(0) \ver 0\rangle$
and $\langle \gamma(q)\vel \bar{q}(x) \Gamma_i G_{\mu\nu}(vx)q(0) \ver 0\rangle$,
representing the non-perturbative contributions. 
These matrix elements can be expressed in terms 
of photon distribution amplitudes (DAs) and wave functions with definite
twists, whose expressions are given in Appendix A. Besides these matrix elements, non-local operators such as two gluons ($\bar q G G q$) and  four quarks ($\bar qq \bar q q$) are expected to seem. However, it is known that the effects of such operators are small, which is justified by the conformal spin expansion \cite{Balitsky:1987bk,Braun:1989iv}, and thus we shall ignore them.
%
The QCD representation of the correlation function is obtained by using Eqs.~(\ref {QCDside1}-\ref {QCDES5}).  Then, the Fourier transformation
is applied to transfer expressions in x-space to the momentum space.

To find the desired sum rules, we obtain the invariant amplitude $\Pi^{QCD}(p,q)$ corresponding to the structure $\eslash \qslash$, and match it to $\Pi^{Had}(p,q)$. We perform the double Borel transformation to both representations of the acquired equality, which is needed to suppress contributions of the higher resonances and continuum states. The last operation to be applied is continuum subtraction, which is obtained by invoking assumption on quark-hadron duality. After these steps, we acquire the required sum rules for the magnetic moments:
\begin{align}
\label{edmn15}
\mu_{P^{1/2}_{QQ}} \,\lambda^2_{P^{1/2}_{QQ}}\, m_{P^{1/2}_{QQ}} =e^{\frac{m^2_{P^{1/2}_{QQ}}}{M^2}}\, \Delta^{QCD}.
\end{align}
%
Explicit forms of the analytical expressions obtained for the $\Delta^{QCD}$ function are given in  Appendix B.

\subsection{Formalism of the \texorpdfstring{$P^{3/2}_{QQ}$}{} states}

In the present subsection, we derive the light-cone sum rule for the magnetic moments of the  $P^{3/2}_{QQ}$ pentaquark states. To do this, we begin with subsequent correlation function,

\begin{equation}
 \label{Pc101}
\Pi _{\mu \nu }(p,q)=i\int d^{4}xe^{ip\cdot x}\langle 0|\mathcal{T}\{J_{\mu}^{P^{3/2}_{QQ}}(x)
\bar J_{\nu }^{P^{3/2}_{QQ}}(0)\}|0\rangle_{\gamma}, 
\end{equation}%
where the interpolating current of doubly-heavy pentaquark states with  $J^P = \frac{3}{2}^{-}$ is denoted $J_{\mu(\nu)}^{P^{3/2}_{QQ}}$. In the compact pentaquark  picture, it is given as~\cite{Wang:2018lhz}

\begin{eqnarray}
 J_\mu^{P^{3/2}_{QQ}}(x)&=& \varepsilon^{abc}\varepsilon^{ade} \varepsilon^{bfg} \big[Q^T_d(x) C\gamma_\mu Q_e(x)\big] \big[u^T_f(x) C\gamma_5 d_g(x)\big]   C \bar{q}^T_c(x) \, ,
\end{eqnarray}

The correlation function obtained depending on the hadron parameters is written as,
\begin{eqnarray}\label{Pc103}
\Pi^{Had}_{\mu\nu}(p,q)&=&\frac{\langle0\mid  J_{\mu}^{P^{3/2}_{QQ}}(x)\mid
{P^{3/2}_{QQ}}(p,s)\rangle}{[p^{2}-m_{{P^{3/2}_{QQ}}}^{2}]}\langle {P^{3/2}_{QQ}}(p,s)\mid
{P^{3/2}_{QQ}}(p+q,s)\rangle_\gamma 
\frac{\langle {P^{3/2}_{QQ}}(p+q,s)\mid
\bar{J}_{\nu}^{P^{3/2}_{QQ}}(0)\mid 0\rangle}{[(p+q)^{2}-m_{{P^{3/2}_{QQ}}}^{2}]}+...
\end{eqnarray}
The matrix element of the interpolating current 
between the vacuum and the $P^{3/2}_{QQ}$ pentaquark is defined as
\begin{align}\label{lambdabey}
\langle0\mid J_{\mu}^{P^{3/2}_{QQ}}(x)\mid {P^{3/2}_{QQ}}(p,s)\rangle&=\lambda_{{P^{3/2}_{QQ}}}u_{\mu}(p,s),\nonumber\\
\langle {P^{3/2}_{QQ}}(p+q,s)\mid
\bar{J}_{\nu}^{P^{3/2}_{QQ}}(0)\mid 0\rangle &= \lambda_{{P^{3/2}_{QQ}}}\bar u_{\nu}(p+q,s), 
\end{align}
where the $u_{\mu}(p,s)$, $u_{\nu}(p+q,s)$ and $\lambda_{{P^{3/2}_{QQ}}}$ are the spinors and residue doubly-heavy $P^{3/2}_{QQ}$ pentaquark states, respectively. 

The transition matrix element $\langle
{P^{3/2}_{QQ}}(p)\mid {P^{3/2}_{QQ}}(p+q)\rangle_\gamma$ entering Eq.
(\ref{Pc103}) can be written as follows
\cite{Weber:1978dh,Nozawa:1990gt,Pascalutsa:2006up,Ramalho:2009vc}:
\begin{align}\label{matelpar}
\langle {P^{3/2}_{QQ}}(p,s)\mid {P^{3/2}_{QQ}}(p+q,s)\rangle_\gamma &=-e\bar
u_{\mu}(p,s)\Bigg[F_{1}(q^2)g_{\mu\nu}\eslash-
\frac{1}{2m_{{P^{3/2}_{QQ}}}} 
\Big[F_{2}(q^2)g_{\mu\nu}+F_{4}(q^2)\frac{q_{\mu}q_{\nu}}{(2m_{{P^{3/2}_{QQ}}})^2}\Big]\eslash\qslash
\nonumber\\&+
F_{3}(q^2)\frac{1}{(2m_{{P^{3/2}_{QQ}}})^2}q_{\mu}q_{\nu}\eslash \Bigg] u_{\nu}(p+q,s).
\end{align}
where $F_i$'s are the Lorentz invariant form factors. 

In principle, we can derive the hadronic representation of the correlation function employing Eqs. (\ref{Pc101})-(\ref{matelpar}), but in this case we run into two undesirable problems. The first of these problems is that the Lorentz structures in the correlation function are not independent, and the second is that the correlation function also contains spin-1/2 contributions. Indeed, the matrix element 
of the current $J_{\mu}$ between vacuum and spin-1/2 doubly-heavy pentaquarks is non-zero and is determined as
\begin{equation}\label{spin12}
\langle0\mid J_{\mu}(0)\mid B(p,s=1/2)\rangle=(A  p_{\mu}+B\gamma_{\mu})u(p,s=1/2).
\end{equation}
As is seen the undesired spin-1/2 effects are proportional to $\gamma_\mu$ and $p_\mu$.
 By multiplying both sides with $\gamma^\mu$ and using 
 the condition $\gamma^\mu J_\mu = 0$ one can specify the constant A in terms of B.
To eliminate unwanted effects of the spin-1/2 states and acquire only independent Lorentz structures in the correlation function, we carry out the ordering for Dirac
matrices as $\gamma_{\mu}\pslash\eslash\qslash\gamma_{\nu}$ and eliminate expressions 
with $\gamma_\mu$ at the beginning, $\gamma_\nu$ at the end and those proportional to $p_\mu$ and 
$p_\nu$~\cite{Belyaev:1982cd}. Consequently, employing Eqs. (\ref{Pc101})-(\ref{matelpar})
the hadronic side takes the form,
\begin{align}\label{final phenpart}
\Pi^{Had}_{\mu\nu}(p,q)&=\frac{\lambda_{_{{P^{3/2}_{QQ}}}}^{2}}{[(p+q)^{2}-m_{_{{P^{3/2}_{QQ}}}}^{2}][p^{2}-m_{_{{P^{3/2}_{QQ}}}}^{2}]} 
\Bigg[  g_{\mu\nu}\pslash\eslash\qslash \,F_{1}(q^2) 
-m_{{P^{3/2}_{QQ}}}g_{\mu\nu}\eslash\qslash\,F_{2}(q^2)
-
\frac{F_{3}(q^2)}{4m_{{P^{3/2}_{QQ}}}}q_{\mu}q_{\nu}\eslash\qslash\,\nonumber\\
&
-
\frac{F_{4}(q^2)}{4m_{{P^{3/2}_{QQ}}}^3}(\varepsilon.p)q_{\mu}q_{\nu}\pslash\qslash 
+
...
\Bigg].
\end{align}
The final form of the hadronic description associated with the chosen structures as follows:

\begin{eqnarray}
\Pi^{Had}_{\mu\nu}(p,q)&=&\Pi_1^{Had}g_{\mu\nu}\pslash\eslash\qslash \,
+\Pi_2^{Had}g_{\mu\nu}\eslash\qslash\,+
...,
\end{eqnarray}
where $ \Pi_1^{Had} $ and $ \Pi_2^{Had} $ are functions of the form factors $ F_1(q^2) $ and  $ F_2(q^2) $, respectively; and other independent structures and form factors are denoted by dots.

The magnetic form factor, $G_{M}(q^2)$, is characterized with respect to the form factors $F_{i}(q^2)$ as follows~\cite{Weber:1978dh,Nozawa:1990gt,Pascalutsa:2006up,Ramalho:2009vc}:
\begin{align}
G_{M}(q^2) &= [ F_1(q^2) + F_2(q^2)] ( 1+ \frac{4}{5}
\tau ) -\frac{2}{5} [ F_3(q^2)  
+ 
F_4(q^2)] \tau ( 1 + \tau ), 
\end{align}
  where $\tau
= -\frac{q^2}{4m^2_{{P^{3/2}_{QQ}}}}$. At $q^2=0$, the magnetic moment
is obtained  with respect to the functions $F_1(0)$ and $F_2(0)$ form factors as:
\begin{eqnarray}\label{mqo1}
G_{M}(0)&=&F_{1}(0)+F_{2}(0).
\end{eqnarray}
The  magnetic moment, ($\mu_{{P^{3/2}_{QQ}}}$), is described as follows,
 \begin{eqnarray}\label{mqo2}
\mu_{{P^{3/2}_{QQ}}}&=&\frac{e}{2m_{{P^{3/2}_{QQ}}}}G_{M}(0).
\end{eqnarray}

When we perform the above processes, the calculations in terms of hadronic parameters, which are the first step of light-cone sum rule calculations, are completed.

The second step in light-cone sum rule calculations is to evaluate the correlation function in  Eq.~(\ref{Pc101}) in connection with quark-gluon parameters as well as photon DAs. Repeating the processes in the previous subsection gives the subsequent  result:
\begin{eqnarray}
\label{QCDPc11}
\Pi^{QCD}(p,q)&=&i\,\varepsilon_{abc} \varepsilon_{ade} \varepsilon_{bfg} \varepsilon_{a^{\prime}d^{\prime}e^{\prime}} 
\varepsilon_{a^{\prime}b^{\prime}c^{\prime}} \varepsilon_{b^{\prime}f^{\prime}g^{\prime}} \int d^4x \, e^{ip\cdot x} \langle 0\mid  \widetilde S_q^{c^{\prime}c}(-x)
\nonumber\\
&& \Big\{ 
\mbox{Tr}\Big[\gamma_\mu S_{Q}^{ee^\prime}(x) \gamma_\nu \widetilde S_{Q}^{dd^\prime}(x)\Big]\, \mbox{Tr}\Big[\gamma_5 S_d^{gg^\prime}(x)  \gamma_5 \widetilde S_{u}^{ff^\prime}(x)\Big]   \nonumber\\ 
&& 
- 
\mbox{Tr}\Big[\gamma_\mu S_{Q}^{ed^\prime}(x) \gamma_\nu \widetilde S_{Q}^{de^\prime}(x)\Big]\, \mbox{Tr}\Big[\gamma_5 S_d^{gg^\prime}(x)  \gamma_5 \widetilde S_{u}^{ff^\prime}(x)\Big] \Big\}   \mid 0 \rangle _\gamma \, .
\end{eqnarray}

Consequently, the QCD representation of the correlation function in connection with the chosen structures is computed as

\begin{eqnarray}
\Pi^{QCD}_{\mu\nu}(p,q)&=&\Pi_{1}^{QCD}g_{\mu\nu}\pslash\eslash\qslash \,
+\Pi_{2}^{QCD}g_{\mu\nu}\eslash\qslash\,+
....
\end{eqnarray}
Since the $ \Pi_i^{QCD} $  functions are very lengthy, their explicit forms are not given here.

We have obtained the correlation function in terms of both QCD and hadronic parameters. For the magnetic moment calculations, the QCD and hadronic descriptions of the correlation function are equalized using the quark-hadron duality ansatz.
By matching the coefficients of the structures $g_{\mu\nu}\pslash\eslash\qslash$ and $g_{\mu\nu}\eslash\qslash$, respectively for the $F_1$ and  $F_2$ we get light-cone sum rules for these two form factors. Consequently, we acquire,

\begin{eqnarray}
\Pi^{Had}_{\mu\nu}(p,q)= \Pi^{QCD}_{\mu\nu}(p,q).
\end{eqnarray}

%
%
Analytical expressions have also been obtained for the $P_{QQ}^{3/2}$ doubly-heavy pentaquarks. The next step will be to perform numerical calculations for both $P_{QQ}^{1/2}$ and $P_{QQ}^{3/2}$  doubly-heavy pentaquarks.

\end{widetext}

\section{Numerical analysis}\label{numerical}

 The light-cone sum rule for magnetic moments of the $P^{1/2}_{QQ}$ and $P^{3/2}_{QQ}$ states contains many input parameters that we need their numerical values.  
We use  $m_u=m_d=0$, $m_c = 1.275 \pm 0.02\,\mbox{GeV}$, $m_b = 4.18^{+0.03}_{-0.02}\,\mbox{GeV}$~\cite{Patrignani:2016xqp},    $f_{3\gamma}=-0.0039~\mbox{GeV}^2$~\cite{Ball:2002ps}, $\langle \bar uu\rangle = \langle \bar dd\rangle=(-0.24 \pm 0.01)^3\,\mbox{GeV}^3$~\cite{Ioffe:2005ym}, $m_0^{2} = 0.8 \pm 0.1 \,\mbox{GeV}^2$ \cite{Ioffe:2005ym},  
$\langle g_s^2G^2\rangle = 0.88~ \mbox{GeV}^4$~\cite{Matheus:2006xi}.   
To further the numerical analysis we also need the numerical values of the mass and residues of these states.  In Ref. \cite{Wang:2018lhz}, these values were obtained within the framework of the mass sum rules. The obtained results for masses and residues are given as $m_{P_{cc}^{1/2}} = 4.21^{+0.10}_{-0.11}~ \mbox{GeV}$,
$m_{P_{cc}^{3/2}} = 4.27^{+0.11}_{-0.10}~ \mbox{GeV}$, 
$m_{P_{bb}^{1/2}} = 10.75^{+0.12}_{-0.12}~ \mbox{GeV}$, $m_{P_{bb}^{3/2}} = 10.76^{+0.11}_{-0.13}~ \mbox{GeV}$, $\lambda_{P_{cc}^{1/2}} = (2.51^{+0.46}_{-0.39}) \times 10^{-3}~ \mbox{GeV}^6$,
$\lambda_{P_{cc}^{3/2}} = (1.65^{+0.30}_{-0.25}) \times 10^{-3}~ \mbox{GeV}^6$,
$\lambda_{P_{bb}^{1/2}} = (7.53^{+1.52}_{-1.39}) \times 10^{-3}~ \mbox{GeV}^6$ and 
$\lambda_{P_{bb}^{3/2}} = (4.27^{+0.85}_{-0.78}) \times 10^{-3}~ \mbox{GeV}^6$.
One of the fundamental components of the light-cone sum rules for the magnetic moment is the photon DAs. The values of DAs are obtained in Ref.~\cite{Ball:2002ps}, which we will employ in our numerical computations. The explicit expressions of the photon DAs are given in Appendix A.

The light-cone sum rules calculation for magnetic moments of doubly-heavy pentaquarks also contains two arbitrary parameters, the Borel mass $M^2$ and the continuum threshold $s_0$. According to the philosophy of the method used, we should find the working intervals in which the magnetic moments are practically insensitive to variations of these parameters.
To do this, two constraints are applied, such as pole contribution (PC) and convergence of operator product expansion (OPE).
Our numerical analysis indicates that the requirements of the method are fulfilled in the regions of arbitrary parameters presented as
\begin{align}
 23.0~\mbox{GeV}^2 \leq  s_0 \leq 25.0~\mbox{GeV}^2,\nonumber\\
 5.0~\mbox{GeV}^2 \leq M^2 \leq 7.0~\mbox{GeV}^2,
\end{align}
for doubly-charmed pentaquark states and 
\begin{align}
 121.0~\mbox{GeV}^2 \leq  s_0 \leq 125.0~\mbox{GeV}^2,\nonumber\\
 11.0~\mbox{GeV}^2 \leq M^2 \leq 15.0~\mbox{GeV}^2,
\end{align}
for doubly-bottom pentaquark states.  In our calculations, PC changes on average within limits $0.33 \leq$ PC $\leq 0.58$, which is acceptable for multiquark states.  When we examine the OPE convergence, we have obtained that the contribution of the higher dimensional term in OPE is less than $\sim 1 \%$. As can be seen from these results, the chosen working regions for $M^2$ and $s_0$ meet the requirements of the method. 

\begin{widetext}
 
\begin{figure}[htp]
\centering
\subfloat[]{\includegraphics[width=0.45\textwidth]{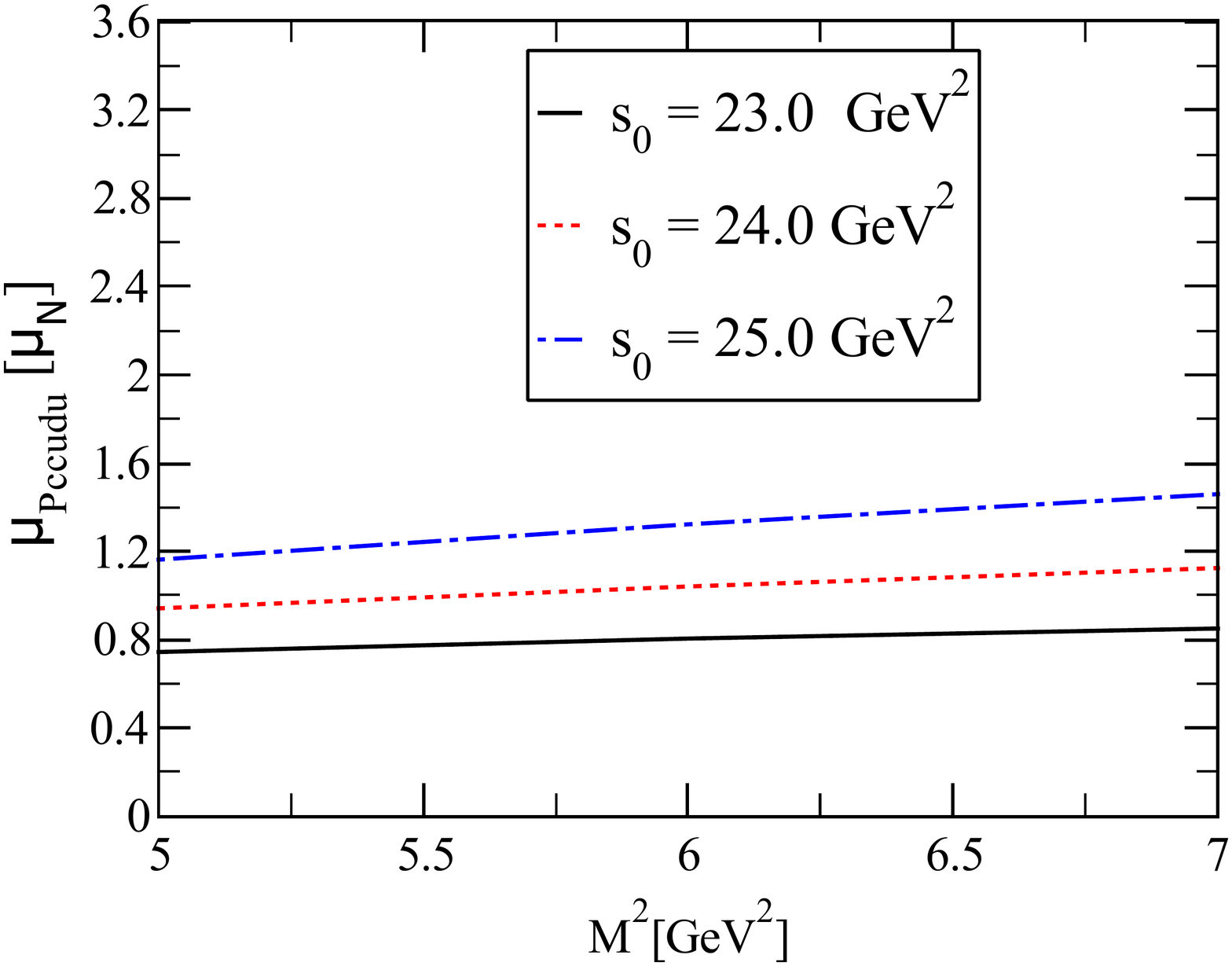}}
\subfloat[]{\includegraphics[width=0.45\textwidth]{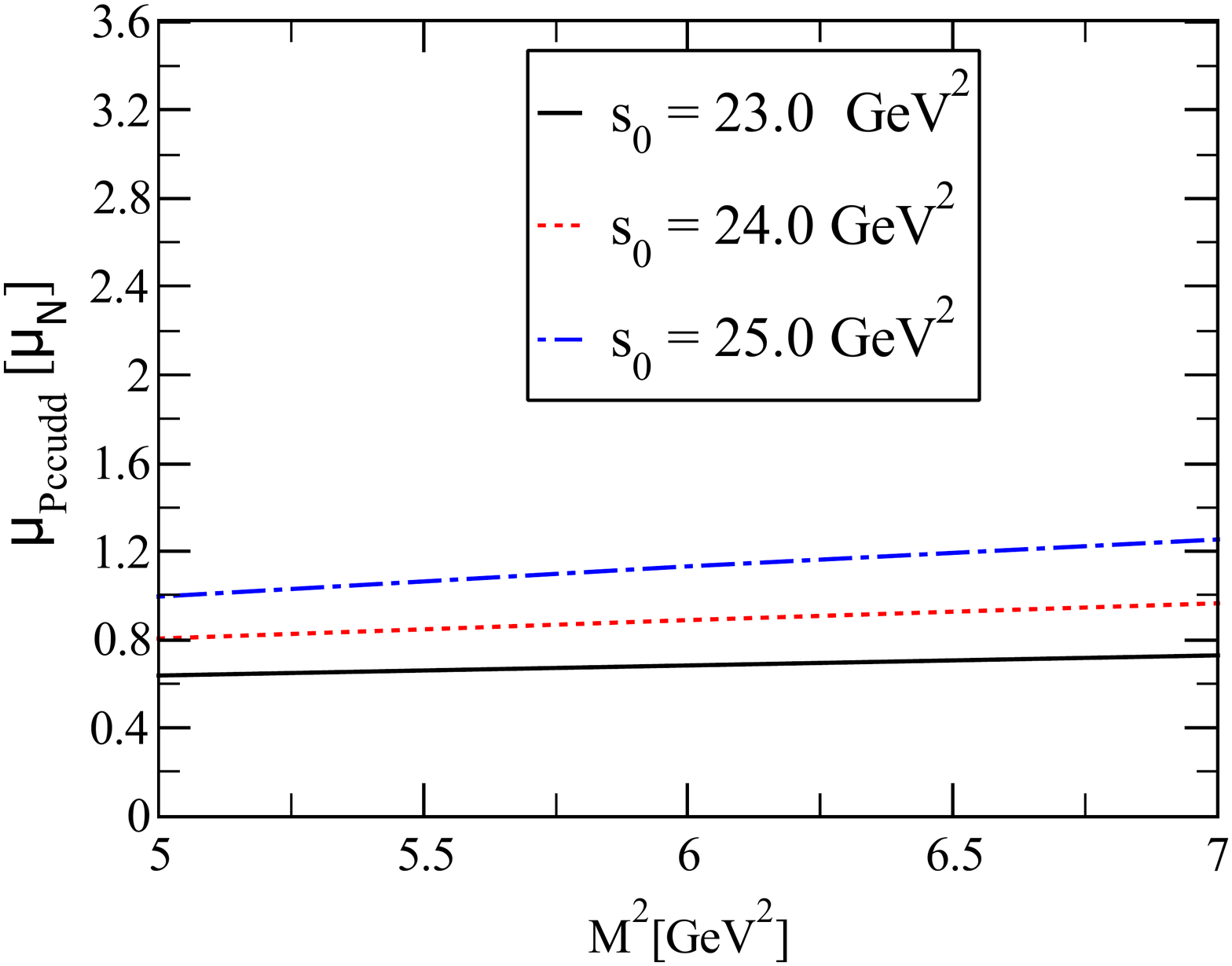}}\\
\subfloat[]{\includegraphics[width=0.45\textwidth]{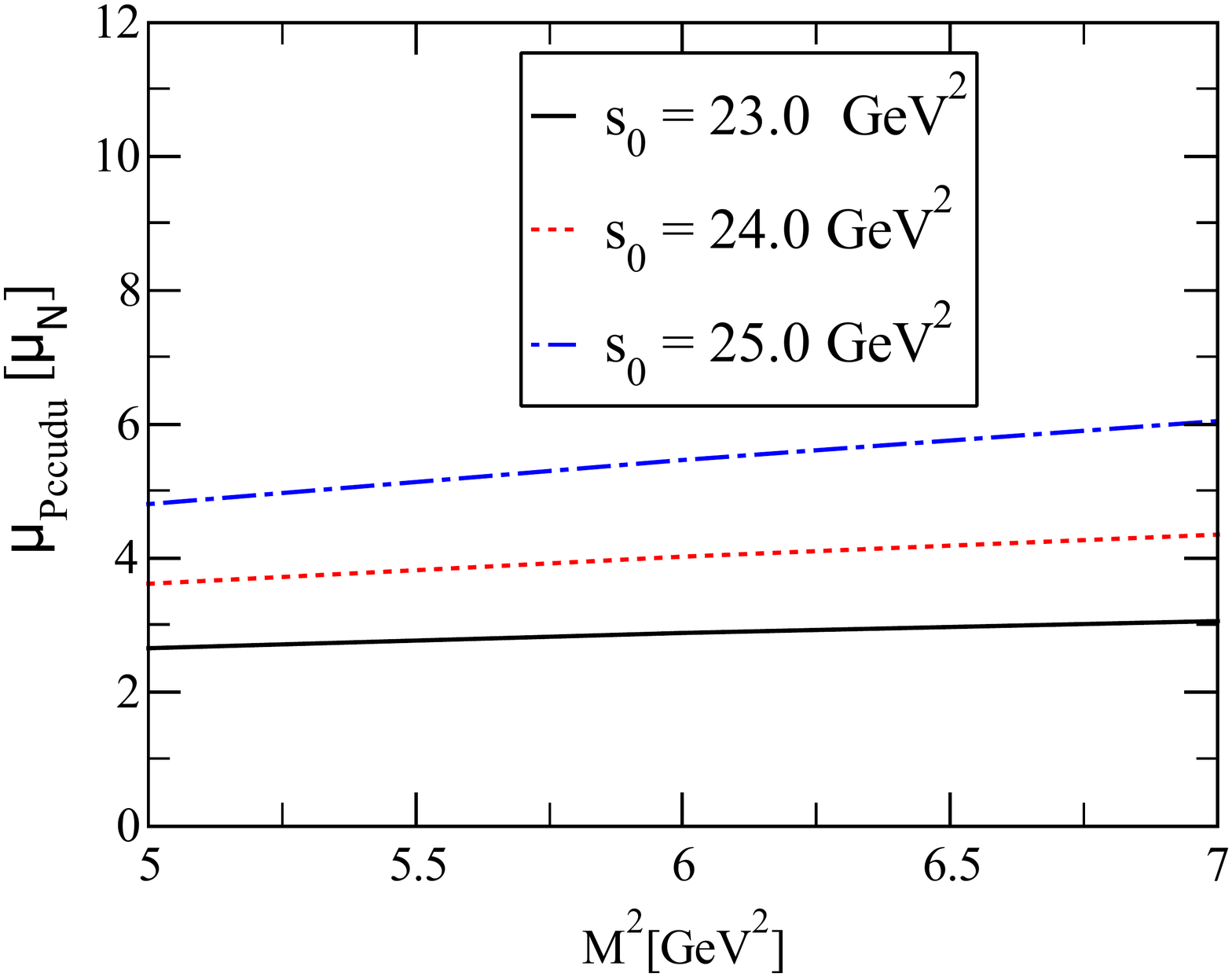}}
\subfloat[]{\includegraphics[width=0.45\textwidth]{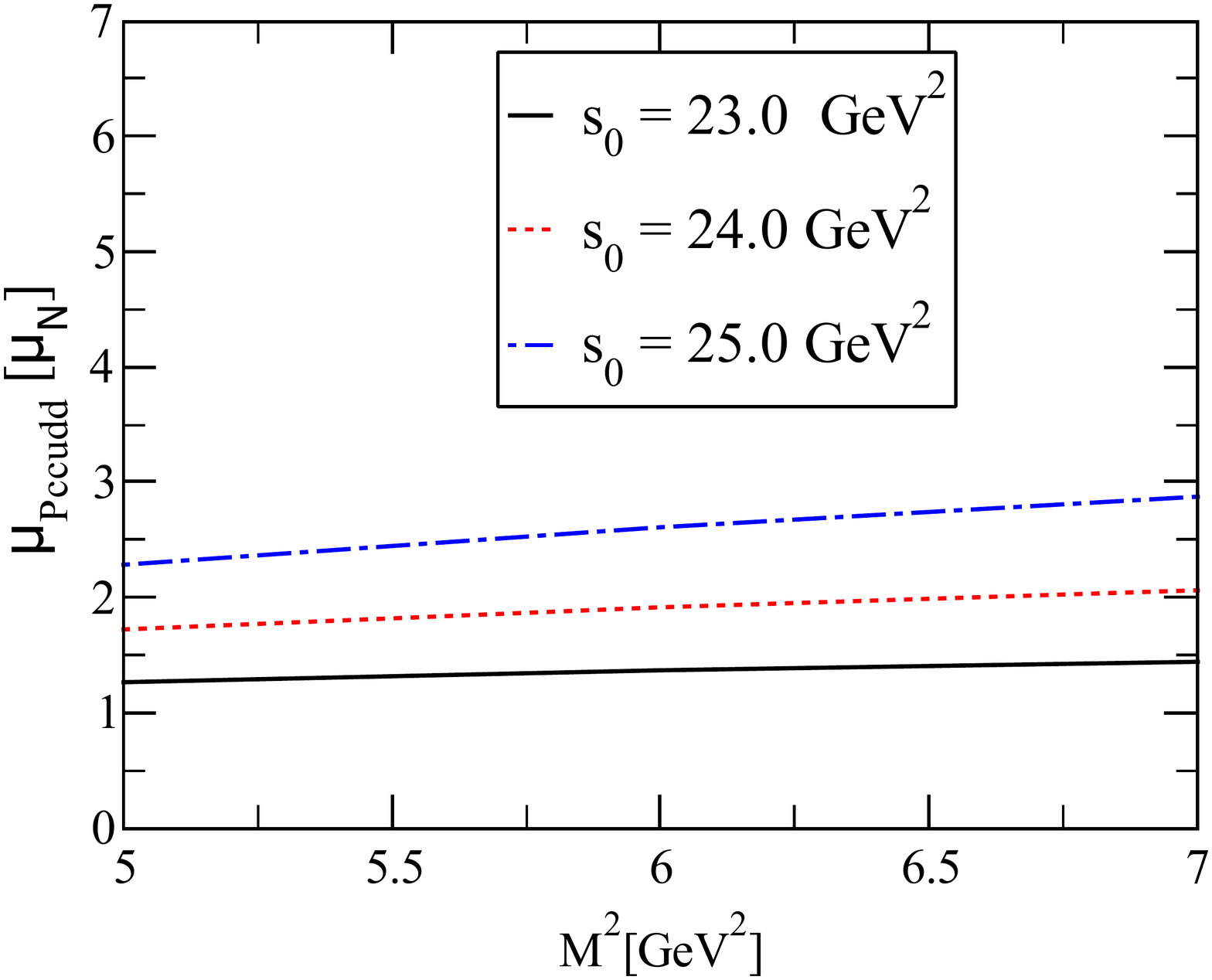}}
 \caption{Dependencies of magnetic moments of  $P^{1/2}_{cc}$ and $P^{3/2}_{cc}$ states on $M^2$ at three different values of $s_0$; (a) and (b) for $P^{1/2}_{cc}$  states; and  (c) and (d) for $P^{3/2}_{cc}$ states.}
 \label{Msqfig}
  \end{figure}
  
  \end{widetext}

In Fig. \ref{Msqfig}, as an example,  we plot the dependencies of the magnetic moments of doubly-charmed pentaquark states on $M^2$ at several fixed values of the $s_0$. As can be seen from the figure, though being not completely insensitive, the magnetic moments show reasonable dependency on the arbitrary parameters, $s_0$ and $M^2$ which is acceptable in the error limits of the light-cone sum rule method.

 We have determined all the parameters we need to complete the numerical analysis of magnetic moments. 
 Using the values of the parameters we determined, we give the numerical results we obtained for the magnetic moment as follows
 \begin{align}\label{pcc12}
 \mu_{P_{ccud\bar u}} &= 1.08^{+0.38}_{-0.34} \,\mu_N, ~~~~~~~  
 \mu_{P_{bbud\bar u}} = -2.48^{+0.74}_{-0.67} \,\mu_N, \\
 \nonumber\\
 \mu_{P_{ccud\bar d}} &= 0.93^{+0.32}_{-0.29} \,\mu_N, ~~~~~~~ 
  \mu_{P_{bbud\bar d}} =- 3.09^{+0.93}_{-0.84} \,\mu_N,
\end{align}
for spin-1/2 doubly-heavy pentaquark states,
\begin{align}
 \mu_{P_{ccud\bar u}} &= 4.23^{+2.00}_{-1.58} \,\mu_N, ~~~~~~~ 
 \mu_{P_{bbud\bar u}} = 5.46^{+1.04}_{-1.00} \,\mu_N, \\
 \nonumber\\
 \mu_{P_{ccud\bar d}} &= 2.01^{+0.86}_{-0.75} \,\mu_N,~~~~~~~ 
 \mu_{P_{bbud\bar d}} = 3.14^{+0.64}_{-0.61} \,\mu_N,\label{pcc32}
\end{align}
for spin-3/2 doubly-heavy pentaquark states.
The errors in the results given in Eqs.~(\ref{pcc12}) and (\ref{pcc32}) are due to all input parameters, extra parameters such as $s_0$ and $M^2$, as well as the parameters on which the wave functions used in the photon distribution amplitudes depend.
The magnetic moments of the doubly-heavy pentaquark states have been extracted from the light-cone sum rules employing for their physical representations a simple-pole approximation [see, Eqs. (\ref{edmn02}) and (\ref{Pc103})]. In the case of the multiquark hadrons such approximation should be verified by supplementary arguments, because a physical side of relevant sum rules receives contributions from two-hadron reducible terms as well. This problem was first proposed during theoretical studies of the pentaquarks \cite{Kondo:2004cr,Lee:2004xk}. Two-hadron contaminating terms have to be considered when extracting parameters of multiquark hadrons. In the case of the multiquark hadrons they lead to modification in the quark propagator
\begin{equation}
\frac{1}{m^{2}-p^{2}}\rightarrow \frac{1}{m^{2}-p^{2}-i\sqrt{p^{2}}\Gamma (p)%
},  \label{eq:Modif}
\end{equation}%
where $\Gamma (p)$\ is the finite width of the  multiquark hadrons generated by two-hadron scattering states.  When these effects are properly considered in the sum rules,  they rescale the residue of the  multiquark hadrons under investigations leaving its mass unchanged. Detailed investigations show that two-hadron scattering effects are small  for  multiquark hadrons (see Refs. \cite{Lee:2004xk,Wang:2015nwa,Agaev:2018vag,Sundu:2018nxt,Wang:2019hyc,Albuquerque:2021tqd,Albuquerque:2020hio,Wang:2020iqt,Wang:2019igl,Wang:2020eew,Pimikov:2019dyr}).
Therefore, in the present study the zero-width single-pole approximation has been used.

 When the results in Eqs.~(\ref{pcc12})-(\ref{pcc32}) are examined, it can be seen that the results acquired for the magnetic moments are of measurable size in the experiments.
 It is seen that the spin-1/2 doubly-heavy pentaquarks results are close to each other, but the difference between the results of the spin-3/2 doubly-heavy pentaquark states is on the order of two.
 To understand the reason for this difference, we extracted the individual quark contributions to the magnetic moments. 
This can be done by dialing the corresponding charge factors $e_q$ and $e_Q$. In case of spin-1/2 doubly-heavy pentaquarks, we obtained that these magnetic   moments are dominantly determined by the heavy-quarks.  
 The situation is the opposite in the spin-3/2 doubly-heavy pentaquarks. In this case, the dominant contribution comes from light-quarks and the contribution of heavy-quarks is negligible. A more detail investigation shows that the smallness of the heavy-quarks contributions are due to an almost exact cancellation of the expressions involving the heavy-quarks, though these expressions are not small themselves.
 It will be interesting and useful to examine the magnetic moments of these doubly-heavy pentaquark states with different theoretical approaches.

 In Refs.~\cite{Ozdem:2021btf,Ozdem:2018qeh}, the magnetic moments of the $P_c(4312)$ and $P_c(4380)$ hidden-charm pentaquark states have been acquired within light-cone sum rules by assuming them as  diquark-diquark-antiquark and molecular configurations. In these studies, the quark content of both $P_c(4312)$ and $P_c(4380)$ hidden-charm pentaquark states are considered as $c\bar c udu$, and the obtained magnetic moment  results for the diquark-diquark-antiquark picture  have been given as $\mu_{P_c(4312 )}  = 0.40 \pm 0.15~ \mu_N $ and $\mu_{P_c(4380 )}  = 1.30 \pm 0.50~ \mu_N $.
 When the comparison is made for the states with the same quark content, it is seen that there is a significant difference between the magnetic moment results obtained for the hidden-charm and doubly-charm pentaquarks.
 Whether the results obtained in this study are consistent or not can be seen by examining the magnetic moments of these possible doubly-heavy pentaquark states with other theoretical models.
 
 For completeness, we have also acquired higher multipole moments, electric quadrupole ($\mathcal{Q}$) and magnetic octupole ($\mathcal{O}$), of the $P^{3/2}_{QQ}$ pentaquark states as
  \begin{align}
 & \mbox{$P_{ccud\bar u}$ state}: \mathcal{Q} = -0.048^{+0.016}_{-0.014}~\mbox{fm$^2$},~~~~~~~
  \mathcal{O} =- 0.022^{+0.008}_{-0.006}~\mbox{fm$^3$},
  \\
  \nonumber\\
 & \mbox{$P_{bbud\bar u}$ state}: \mathcal{Q} = - 0.22^{+0.07}_{-0.07}~\mbox{fm$^2$},~~~~~~~~~~  
  \mathcal{O} = - 0.10^{+0.04}_{-0.03}~\mbox{fm$^3$},
  \\
  \nonumber\\
 & \mbox{$P_{ccud\bar d}$ state}: \mathcal{Q}= 0.024^{+0.008}_{-0.008}~\mbox{fm$^2$},~~~~~~~~~
  \mathcal{O} = 0.011^{+0.005}_{-0.004}~\mbox{fm$^3$},
  \\
  \nonumber\\
 & \mbox{$P_{bbud\bar d}$ state}: \mathcal{Q}=  0.11^{+0.04}_{-0.03}~\mbox{fm$^2$},~~~~~~~~~~~~
  \mathcal{O} = 0.06^{+0.002}_{-0.002}~\mbox{fm$^3$}.
 \end{align}

 As can be seen from these results, the charge distribution of the $P^{3/2}_{QQ}$ pentaquark states is non-spherical. The sign of electric quadrupole moment is positive for $P_{QQud\bar d}$ pentaquark states and negative for $P_{QQud\bar u}$ pentaquark states, which correspond to the prolate and oblate charge distributions, respectively. 
 
\section{Summary}\label{summary}
 
Motivated by the latest discovery of doubly-charmed tetraquark $T^+_{cc}$ by LHCb collaboration, we have studied the magnetic moments  of the possible doubly-heavy pentaquark states with quantum numbers $J^P = 1/2^-$ and $J^P = 3/2^-$ in the framework of the light-cone sum rules method. In the analysis, these possible pentaquark states are considered in diquark-diquark-antiquark structure. 
We have also acquired non-vanishing values for the electric quadrupole and magnetic octupole moments of spin-3/2 doubly-heavy pentaquark states which mean non-spherical charge distribution.
The magnetic dipole and higher multipole moments  of hadrons encode helpful details about the distributions of the charge and magnetization inside the hadrons, which help us to figure out their geometric configurations.
The discovery of the first doubly-charmed tetraquark state gave a new platform for hadron physics. More theoretical and experimental attempts are required for to figure out its fundamental structure and non-perturbative nature of QCD dynamics in this region.
It would be exciting to predict future experimental attempts that will search possible doubly-heavy pentaquark states and test the obtainment from the present analysis.

\textbf{Data Availability Statement:} This manuscript has no associated data or
the data will not be deposited. [Authors’ comment: This is a theoretical
research work, so no additional data are associated with this work.]

  \begin{widetext}

 \appendix
 \section*{Appendix A: Distribution Amplitudes of the photon }
In the present appendix, the matrix elements $\langle \gamma(q)\vel \bar{q}(x) \Gamma_i q(0) \ver 0\rangle$  
and $\langle \gamma(q)\vel \bar{q}(x) \Gamma_i G_{\mu\nu}q(0) \ver 0\rangle$ associated with the photon DAs are presented as follows \cite{Ball:2002ps},
\begin{eqnarray*}
\label{esbs14}
&&\langle \gamma(q) \vert  \bar q(x) \gamma_\mu q(0) \vert 0 \rangle
= e_q f_{3 \gamma} \left(\varepsilon_\mu - q_\mu \frac{\varepsilon
x}{q x} \right) \int_0^1 du e^{i \bar u q x} \psi^v(u)
\nonumber \\
&&\langle \gamma(q) \vert \bar q(x) \gamma_\mu \gamma_5 q(0) \vert 0
\rangle  = - \frac{1}{4} e_q f_{3 \gamma} \epsilon_{\mu \nu \alpha
\beta } \varepsilon^\nu q^\alpha x^\beta \int_0^1 du e^{i \bar u q
x} \psi^a(u)
\nonumber \\
&&\langle \gamma(q) \vert  \bar q(x) \sigma_{\mu \nu} q(0) \vert  0
\rangle  = -i e_q \langle \bar q q \rangle (\varepsilon_\mu q_\nu - \varepsilon_\nu
q_\mu) \int_0^1 du e^{i \bar u qx} \left(\chi \varphi_\gamma(u) +
\frac{x^2}{16} \mathbb{A}  (u) \right) \nonumber \\ 
&&-\frac{i}{2(qx)}  e_q \bar qq \left[x_\nu \left(\varepsilon_\mu - q_\mu
\frac{\varepsilon x}{qx}\right) - x_\mu \left(\varepsilon_\nu -
q_\nu \frac{\varepsilon x}{q x}\right) \right] \int_0^1 du e^{i \bar
u q x} h_\gamma(u)
\nonumber \\
&&\langle \gamma(q) | \bar q(x) g_s G_{\mu \nu} (v x) q(0) \vert 0
\rangle = -i e_q \langle \bar q q \rangle \left(\varepsilon_\mu q_\nu - \varepsilon_\nu
q_\mu \right) \int {\cal D}\alpha_i e^{i (\alpha_{\bar q} + v
\alpha_g) q x} {\cal S}(\alpha_i)
\nonumber \\
&&\langle \gamma(q) | \bar q(x) g_s \tilde G_{\mu \nu}(v
x) i \gamma_5  q(0) \vert 0 \rangle = -i e_q \langle \bar q q \rangle \left(\varepsilon_\mu q_\nu -
\varepsilon_\nu q_\mu \right) \int {\cal D}\alpha_i e^{i
(\alpha_{\bar q} + v \alpha_g) q x} \tilde {\cal S}(\alpha_i)
\nonumber \\
&&\langle \gamma(q) \vert \bar q(x) g_s \tilde G_{\mu \nu}(v x)
\gamma_\alpha \gamma_5 q(0) \vert 0 \rangle = e_q f_{3 \gamma}
q_\alpha (\varepsilon_\mu q_\nu - \varepsilon_\nu q_\mu) \int {\cal
D}\alpha_i e^{i (\alpha_{\bar q} + v \alpha_g) q x} {\cal
A}(\alpha_i)
\nonumber \\
&&\langle \gamma(q) \vert \bar q(x) g_s G_{\mu \nu}(v x) i
\gamma_\alpha q(0) \vert 0 \rangle = e_q f_{3 \gamma} q_\alpha
(\varepsilon_\mu q_\nu - \varepsilon_\nu q_\mu) \int {\cal
D}\alpha_i e^{i (\alpha_{\bar q} + v \alpha_g) q x} {\cal
V}(\alpha_i) \nonumber\\
&& \langle \gamma(q) \vert \bar q(x)
\sigma_{\alpha \beta} g_s G_{\mu \nu}(v x) q(0) \vert 0 \rangle  =
e_q \langle \bar q q \rangle \left\{
        \left[\left(\varepsilon_\mu - q_\mu \frac{\varepsilon x}{q x}\right)\left(g_{\alpha \nu} -
        \frac{1}{qx} (q_\alpha x_\nu + q_\nu x_\alpha)\right) \right. \right. q_\beta
\nonumber \\
 && -
         \left(\varepsilon_\mu - q_\mu \frac{\varepsilon x}{q x}\right)\left(g_{\beta \nu} -
        \frac{1}{qx} (q_\beta x_\nu + q_\nu x_\beta)\right) q_\alpha
-
         \left(\varepsilon_\nu - q_\nu \frac{\varepsilon x}{q x}\right)\left(g_{\alpha \mu} -
        \frac{1}{qx} (q_\alpha x_\mu + q_\mu x_\alpha)\right) q_\beta
\nonumber \\
 &&+
         \left. \left(\varepsilon_\nu - q_\nu \frac{\varepsilon x}{q.x}\right)\left( g_{\beta \mu} -
        \frac{1}{qx} (q_\beta x_\mu + q_\mu x_\beta)\right) q_\alpha \right]
   \int {\cal D}\alpha_i e^{i (\alpha_{\bar q} + v \alpha_g) qx} {\cal T}_1(\alpha_i)
\nonumber \\
 &&+
        \left[\left(\varepsilon_\alpha - q_\alpha \frac{\varepsilon x}{qx}\right)
        \left(g_{\mu \beta} - \frac{1}{qx}(q_\mu x_\beta + q_\beta x_\mu)\right) \right. q_\nu
\nonumber \\ &&-
         \left(\varepsilon_\alpha - q_\alpha \frac{\varepsilon x}{qx}\right)
        \left(g_{\nu \beta} - \frac{1}{qx}(q_\nu x_\beta + q_\beta x_\nu)\right)  q_\mu
\nonumber \\ && -
         \left(\varepsilon_\beta - q_\beta \frac{\varepsilon x}{qx}\right)
        \left(g_{\mu \alpha} - \frac{1}{qx}(q_\mu x_\alpha + q_\alpha x_\mu)\right) q_\nu
\nonumber \\ &&+
         \left. \left(\varepsilon_\beta - q_\beta \frac{\varepsilon x}{qx}\right)
        \left(g_{\nu \alpha} - \frac{1}{qx}(q_\nu x_\alpha + q_\alpha x_\nu) \right) q_\mu
        \right]      
    \int {\cal D} \alpha_i e^{i (\alpha_{\bar q} + v \alpha_g) qx} {\cal T}_2(\alpha_i)
\nonumber \\
  \end{eqnarray*}
\begin{eqnarray*}
&&+\frac{1}{qx} (q_\mu x_\nu - q_\nu x_\mu)
        (\varepsilon_\alpha q_\beta - \varepsilon_\beta q_\alpha)
    \int {\cal D} \alpha_i e^{i (\alpha_{\bar q} + v \alpha_g) qx} {\cal T}_3(\alpha_i)
\nonumber \\ &&+
        \left. \frac{1}{qx} (q_\alpha x_\beta - q_\beta x_\alpha)
        (\varepsilon_\mu q_\nu - \varepsilon_\nu q_\mu)
    \int {\cal D} \alpha_i e^{i (\alpha_{\bar q} + v \alpha_g) qx} {\cal T}_4(\alpha_i)
                        \right\}~,
\end{eqnarray*}
where $\varphi_\gamma(u)$ is the DA of leading twist-2, $\psi^v(u)$,
$\psi^a(u)$, ${\cal A}(\alpha_i)$ and ${\cal V}(\alpha_i)$, are the twist-3 amplitudes, and
$h_\gamma(u)$, $\mathbb{A}(u)$, ${\cal S}(\alpha_i)$, ${\cal{\tilde S}}(\alpha_i)$, ${\cal T}_1(\alpha_i)$, ${\cal T}_2(\alpha_i)$, ${\cal T}_3(\alpha_i)$ 
and ${\cal T}_4(\alpha_i)$ are the
twist-4 photon DAs.
The measure ${\cal D} \alpha_i$ is defined as
\begin{eqnarray*}
\label{nolabel05}
\int {\cal D} \alpha_i = \int_0^1 d \alpha_{\bar q} \int_0^1 d
\alpha_q \int_0^1 d \alpha_g \delta(1-\alpha_{\bar
q}-\alpha_q-\alpha_g)~.\nonumber
\end{eqnarray*}

The expressions of the DAs that entering into the matrix elements above are described as follows:
\begin{eqnarray}
\varphi_\gamma(u) &=& 6 u \bar u \left( 1 + \varphi_2(\mu)
C_2^{\frac{3}{2}}(u - \bar u) \right),
\nonumber \\
\psi^v(u) &=& 3 \left(3 (2 u - 1)^2 -1 \right)+\frac{3}{64} \left(15
w^V_\gamma - 5 w^A_\gamma\right)
                        \left(3 - 30 (2 u - 1)^2 + 35 (2 u -1)^4
                        \right),
\nonumber \\
\psi^a(u) &=& \left(1- (2 u -1)^2\right)\left(5 (2 u -1)^2 -1\right)
\frac{5}{2}
    \left(1 + \frac{9}{16} w^V_\gamma - \frac{3}{16} w^A_\gamma
    \right),
\nonumber \\
h_\gamma(u) &=& - 10 \left(1 + 2 \kappa^+\right) C_2^{\frac{1}{2}}(u
- \bar u),
\nonumber \\
\mathbb{A}(u) &=& 40 u^2 \bar u^2 \left(3 \kappa - \kappa^+
+1\right)  +
        8 (\zeta_2^+ - 3 \zeta_2) \left[u \bar u (2 + 13 u \bar u) \right.
\nonumber \\ && + \left.
                2 u^3 (10 -15 u + 6 u^2) \ln(u) + 2 \bar u^3 (10 - 15 \bar u + 6 \bar u^2)
        \ln(\bar u) \right],
\nonumber \\
{\cal A}(\alpha_i) &=& 360 \alpha_q \alpha_{\bar q} \alpha_g^2
        \left(1 + w^A_\gamma \frac{1}{2} (7 \alpha_g - 3)\right),
\nonumber \\
{\cal V}(\alpha_i) &=& 540 w^V_\gamma (\alpha_q - \alpha_{\bar q})
\alpha_q \alpha_{\bar q}
                \alpha_g^2,
\nonumber \\
{\cal T}_1(\alpha_i) &=& -120 (3 \zeta_2 + \zeta_2^+)(\alpha_{\bar
q} - \alpha_q)
        \alpha_{\bar q} \alpha_q \alpha_g,
\nonumber \\
{\cal T}_2(\alpha_i) &=& 30 \alpha_g^2 (\alpha_{\bar q} - \alpha_q)
    \left((\kappa - \kappa^+) + (\zeta_1 - \zeta_1^+)(1 - 2\alpha_g) +
    \zeta_2 (3 - 4 \alpha_g)\right),
\nonumber \\
{\cal T}_3(\alpha_i) &=& - 120 (3 \zeta_2 - \zeta_2^+)(\alpha_{\bar
q} -\alpha_q)
        \alpha_{\bar q} \alpha_q \alpha_g,
\nonumber \\
{\cal T}_4(\alpha_i) &=& 30 \alpha_g^2 (\alpha_{\bar q} - \alpha_q)
    \left((\kappa + \kappa^+) + (\zeta_1 + \zeta_1^+)(1 - 2\alpha_g) +
    \zeta_2 (3 - 4 \alpha_g)\right),\nonumber \\
{\cal S}(\alpha_i) &=& 30\alpha_g^2\{(\kappa +
\kappa^+)(1-\alpha_g)+(\zeta_1 + \zeta_1^+)(1 - \alpha_g)(1 -
2\alpha_g)\nonumber +\zeta_2[3 (\alpha_{\bar q} - \alpha_q)^2-\alpha_g(1 - \alpha_g)]\},\nonumber \\
\tilde {\cal S}(\alpha_i) &=&-30\alpha_g^2\{(\kappa -\kappa^+)(1-\alpha_g)+(\zeta_1 - \zeta_1^+)(1 - \alpha_g)(1 -
2\alpha_g)\nonumber +\zeta_2 [3 (\alpha_{\bar q} -\alpha_q)^2-\alpha_g(1 - \alpha_g)]\}.
\end{eqnarray}

Numerical values of parameters used in DAs are: $\varphi_2(1~GeV) = 0$, 
$w^V_\gamma = 3.8 \pm 1.8$, $w^A_\gamma = -2.1 \pm 1.0$, $\kappa = 0.2$, $\kappa^+ = 0$, $\zeta_1 = 0.4$, $\zeta_2 = 0.3$.
 
 \end{widetext}

  \begin{widetext}
  \section*{Appendix B: The explicit expression of \texorpdfstring{$\Delta^{QCD} $}{}  function}
 In here, we present the explicit expression for the function $\Delta^{QCD}$ acquired from the light-cone sum rule in subsection \ref{for:Pcc12}. It is acquired by selecting the $\eslash\pslash$ structure as follows
  \begin{align}
  \Delta^{QCD}&=\frac{P_2}{849346560 \,\pi^5}\Bigg \{
  10 P_1 (e_d + e_u)  \Bigg[
   4 m_Q^2 \Bigg (3 m_ 0^2 \Big (I[0, 2, 1, 0] - 2 I[0, 2, 1, 1] + 
          I[0, 2, 1, 2] - 2 I[0, 2, 2, 0]\nonumber\\
          & + 2 I[0, 2, 2, 1]+ 
          I[0, 2, 3, 0] - 
          2 \big (I[1, 1, 1, 0] - 2 I[1, 1, 1, 1] + I[1, 1, 1, 2] - 
              2 I[1, 1, 2, 0] + 2 I[1, 1, 2, 1]  \nonumber\\
              &+ 
              I[1, 1, 3, 0] \big)\Big)+ 
       4 \Big (I[0, 3, 1, 0] - 3 I[0, 3, 1, 1] + 3 I[0, 3, 1, 2] - 
           I[0, 3, 1, 3] - 2 I[0, 3, 2, 0] + 4 I[0, 3, 2, 1]  \nonumber\\
           &+ I[0, 3, 3, 0] - I[0, 3, 3, 1] + 
           3 \big (I[1, 2, 1, 1] - 2 I[1, 2, 1, 2] + I[1, 2, 1, 3] - 
               2 I[1, 2, 2, 1] + 2 I[1, 2, 2, 2]\nonumber\\
               &- 
           2 I[0, 3, 2, 2]+ 
               I[1, 2, 3, 1]\big)\Big)\Bigg) + 
    3  \Big (I[0, 4, 2, 0] - 3 I[0, 4, 2, 1] + 3 I[0, 4, 2, 2] - 
        I[0, 4, 2, 3] - 3 I[0, 4, 3, 0]  \nonumber\\
        &+ 6 I[0, 4, 3, 1]- 
        3 I[0, 4, 3, 2] + 3 I[0, 4, 4, 0] - 3 I[0, 4, 4, 1] - 
        I[0, 4, 5, 0] + 
        4 \big (3 I[1, 3, 2, 1] - 3 I[1, 3, 2, 2] \nonumber\\
        & + I[1, 3, 2, 3]+ 
            3 \big (-2 I[1, 3, 3, 1] + I[1, 3, 3, 2] + 
                I[1, 3, 4, 1]\big)\big)\Big)\Bigg]\nonumber
                \end{align}
               \begin{align}
     &+9 e_Q \Bigg[
   16 m_Q^2 \Bigg (5 P_ 1 \Big (I[0, 3, 1, 0] - 2 I[0, 3, 1, 1] + 
          I[0, 3, 1, 2] - 2 I[0, 3, 2, 0] + 2 I[0, 3, 2, 1] + 
          I[0, 3, 3, 0]\Big) \nonumber\\
          &- 
       45 m_ 0^2 \Big (I[0, 4, 1, 1] - 2 I[0, 4, 1, 2] + 
          I[0, 4, 1, 3] - 2 I[0, 4, 2, 1] + 2 I[0, 4, 2, 2] + 
          I[0, 4, 3, 1]\Big) + 
       18\Big (I[0, 5, 1, 2] \nonumber\\
       &- 2 I[0, 5, 1, 3] + I[0, 5, 1, 4] - 
           2 I[0, 5, 2, 2] + 2 I[0, 5, 2, 3] + 
           I[0, 5, 3, 2]\Big)\Bigg) + 
    3  \Bigg (5 P_ 1 \Big (I[0, 4, 2, 0] - 3 I[0, 4, 2, 1] \nonumber\\
    &+ 
          3 I[0, 4, 2, 2] - I[0, 4, 2, 3] - 3 I[0, 4, 3, 0] + 
          6 I[0, 4, 3, 1] - 3 I[0, 4, 3, 2] + 3 I[0, 4, 4, 0] - 
          3 I[0, 4, 4, 1] - I[0, 4, 5, 0] \nonumber\\
          &+ 
          4 \big (3 I[1, 3, 2, 1] - 3 I[1, 3, 2, 2] + I[1, 3, 2, 3] + 
              3 \big (-2 I[1, 3, 3, 1] + I[1, 3, 3, 2] + 
                  I[1, 3, 4, 1]\big)\big)\Big) \nonumber\\
                  &+ 
       36 m_ 0^2\Big (I[0, 5, 2, 0] - 4 I[0, 5, 2, 1] + 
           6 I[0, 5, 2, 2] - 4 I[0, 5, 2, 3] + I[0, 5, 2, 4] - 
           3 I[0, 5, 3, 0] + 9 I[0, 5, 3, 1] \nonumber\\
           &- 9 I[0, 5, 3, 2] + 
           3 I[0, 5, 3, 3] + 3 I[0, 5, 4, 0] - 6 I[0, 5, 4, 1] + 
           3 I[0, 5, 4, 2] - I[0, 5, 5, 0] + I[0, 5, 5, 1] \nonumber\\
           &+ 
           5 \big (I[1, 4, 2, 1] - 3 I[1, 4, 2, 2] + 3 I[1, 4, 2, 3] -
                I[1, 4, 2, 4] - 
               3 \big (I[1, 4, 3, 1] - 2 I[1, 4, 3, 2] + 
                  I[1, 4, 3, 3] \nonumber\\
                  &- I[1, 4, 4, 1] + I[1, 4, 4, 2]\big) -
                I[1, 4, 5, 1]\big)\Big)\Bigg) - 
    36  \Bigg (2 I[0, 6, 2, 1] - 7 I[0, 6, 2, 2] + 9 I[0, 6, 2, 3] - 
        5 I[0, 6, 2, 4] \nonumber\\
        &+ I[0, 6, 2, 5] - 6 I[0, 6, 3, 1] + 
        15 I[0, 6, 3, 2] - 12 I[0, 6, 3, 3] + 3 I[0, 6, 3, 4] + 
        6 I[0, 6, 4, 1] - 9 I[0, 6, 4, 2]\nonumber\\
        &+ 3 I[0, 6, 4, 3] - 
        2 I[0, 6, 5, 1] + I[0, 6, 5, 2] + 
        6 \Big (I[1, 5, 2, 2] - 3 I[1, 5, 2, 3] + 3 I[1, 5, 2, 4] - 
            I[1, 5, 2, 5] \nonumber\\
            &- 
            3 \big (I[1, 5, 3, 2] - 2 I[1, 5, 3, 3] + I[1, 5, 3, 4] - 
               I[1, 5, 4, 2] + I[1, 5, 4, 3]\big) - 
            I[1, 5, 5, 2]\Big)\Bigg)\Bigg]    \Bigg\}   \nonumber\\
   &+\frac{P_2}{(13589544960 \, \pi^5}\Bigg\{
   2560 (e_d - 
   e_u) f_ {3\gamma} m_Q^2 P_ 1 \pi^2 (2 I_ 2[\mathcal {A}] I[
     0, 2, 2, 0] - I_ 5[\psi_a] I[0, 2, 3, 0])\nonumber\\
     &
     -80 m_Q^2 \Big (144 (2 e_d + 
      e_u) f_ {3\gamma} m_ 0^2\pi^2 I_ 2[\mathcal {V}] + 
   e_q P_ 1 \big (-44 I_1[\mathcal {S}] - 33 I_ 1[\mathcal T_2] - 
       184 I_3[\mathcal {S}] + 138 I_ 3[\mathcal T_1] + 
       193 I_3[\mathcal T_2]\nonumber\\
       &+ 
       11 I_3[\mathcal T_4]\big)\Big) I[0, 3, 3, 0]
       -345 e_q P_ 1 \Big (4 I_ 1[\mathcal S] - 3 I_ 1[\mathcal T_ 2] + 
    6 I_ 3[\mathcal T_ 2]\Big) I[0, 4, 5, 0] + 
 360 (2 e_d + 
    e_u) f_ {3\gamma} \pi^2 I_ 2[\mathcal V]\nonumber\\
    &\times \Big (32 m_Q^2 I[0, 4, 3,
       0] + 3 m0^2 I[0, 4, 5, 0]\Big)
       +864 \Big (8 e_q m_Q^2 \big (4 I_3[\mathcal S] - 3 I_3[\mathcal T_1] - 
      3 I_3[\mathcal T_2]\big) I[0, 5, 3, 
     0] \nonumber\\
     &+ (2 e_d + e_u) f_ {3\gamma} \pi^2 I_ 2[\mathcal V] I[0, 5, 5,
       0]\Big)
       -432  e_q \Big (4 I_ 1[\mathcal S] - 3 I_ 1[\mathcal T_ 2] + 
   2 (-8 I_ 3[\mathcal S] + 6 I_ 3[\mathcal T_ 1] + 
       8 I_ 3[\mathcal T_ 2] + I_ 3[\mathcal T_ 4])\Big)\nonumber\\
       &\times I[0, 6, 5, 0]\nonumber\\
       &+2560 (e_d - 
   e_u)  f_ {3\gamma} m_Q^2 P_ 1 \pi^2 \Big (I[0, 2, 1, 0] - 
    2 I[0, 2, 1, 1] + I[0, 2, 1, 2] - 2 I[0, 2, 2, 0] + 
    2 I[0, 2, 2, 1] + I[0, 2, 3, 0]\Big) \nonumber\\
    &\times \psi_a[u_0]\Bigg\},
    \label{appson}
  \end{align}
where $P_1 =\langle g_s^2 G^2\rangle$ is gluon condensate, $P_2 =\langle \bar q q \rangle$ stands for u/d quark condensate. We should also remark that in the Eq.(\ref{appson}), for simplicity we have only given the terms that give significant contributions to the numerical values of the magnetic moments and neglected to give many higher dimensional operators though they have been taken into account in the numerical analyses.
  The~$I[n,m,l,k]$, $I_1[\mathcal{F}]$, ~$I_2[\mathcal{F}]$, ~$I_3[\mathcal{F}]$, ~$I_4[\mathcal{F}]$, 
~$I_5[\mathcal{F}]$, and ~$I_6[\mathcal{F}]$ functions are
defined as:
\begin{align}
 I[n,m,l,k]&= \int_{4 m_Q^2}^{s_0} ds \int_{0}^1 dt \int_{0}^1 dw~ e^{-s/M^2}~
 s^n\,(s-4\,m_Q^2)^m\,t^l\,w^k,\nonumber\\
 I_1[\mathcal{F}]&=\int D_{\alpha_i} \int_0^1 dv~ \mathcal{A}(\alpha_{\bar q},\alpha_q,\alpha_g)
 \delta'(\alpha_ q +\bar v \alpha_g-u_0),\nonumber\\
  I_2[\mathcal{F}]&=\int D_{\alpha_i} \int_0^1 dv~ \mathcal{A}(\alpha_{\bar q},\alpha_q,\alpha_g)
 \delta'(\alpha_{\bar q}+ v \alpha_g-u_0),\nonumber\\
    I_3[\mathcal{F}]&=\int D_{\alpha_i} \int_0^1 dv~ \mathcal{A}(\alpha_{\bar q},\alpha_q,\alpha_g)
 \delta(\alpha_ q +\bar v \alpha_g-u_0),\nonumber
   \end{align}
 \begin{align}
   I_4[\mathcal{F}]&=\int D_{\alpha_i} \int_0^1 dv~ \mathcal{A}(\alpha_{\bar q},\alpha_q,\alpha_g)
 \delta(\alpha_{\bar q}+ v \alpha_g-u_0),\nonumber\\
   I_5[\mathcal{F}]&=\int_0^1 du~ A(u)\delta'(u-u_0),\nonumber\\
 I_6[\mathcal{F}]&=\int_0^1 du~ A(u),\nonumber
 \end{align}
 where $\mathcal{F}$ denotes the corresponding photon DAs.
 
\end{widetext}

\bibliography{PQQMM}
\end{document}